# A University of Texas Medical Branch Case Study on Aortic Calcification Detection

Eric Walser[1], MD; Peter McCaffrey[2], MD, MS, FCAP, Chief AI Officer; Kal Clark[3,4], MD; Nicholas Czarnek[4], PhD

[1]Department of Radiology, The University of Texas Medical Branch, Galveston, TX, USA

[2]Department of Pathology, The University of Texas Medical Branch, Galveston, TX, USA

[3]Department of Radiology, The University of Texas Health San Antonio, San Antonio, TX, USA

[4]Zauron Labs, Inc.

## Abstract

This case study details The University of Texas Medical Branch (UTMB)'s partnership with Zauron Labs, Inc. to enhance detection and coding of aortic calcifications (ACs) using chest radiographs. ACs are often underreported despite their significant prognostic value for cardiovascular disease, and UTMB partnered with Zauron to apply its advanced AI tools, including a high-performing image model (AUC = 0.938) and a fine-tuned language model based on Meta's Llama 3.2, to retrospectively analyze imaging and report data. The effort identified 495 patients out of 3,988 unique patients assessed (5,000 total exams) whose reports contained indications of aortic calcifications that were not properly coded for reimbursement (12.4% miscode rate) as well as an additional 84 patients who had aortic calcifications that were missed during initial review (2.1% misdiagnosis rate). Identification of these patients provided UTMB with the potential to impact clinical care for these patients and pursue $314k in missed annual revenue. These findings informed UTMB's decision to adopt Zauron's Guardian Pro software system-wide to ensure accurate, AI-enhanced peer review and coding, improving both patient care and financial solvency. This study is covered under University of Texas Health San Antonio's Institutional Review Board Study ID 00001887.

## Section A - Aortic Calcification Detection Using Chest Radiographs: Implications for Patient Management and Prognosis

Aortic calcifications (ACs), or calcium deposits within the aortic wall layers like the examples shown in Figure 1, are common pathological findings that increase in prevalence with advancing age and frequently coexist with atherosclerosis[1–3]. As the largest artery in the body, the aorta is susceptible to accumulation of calcium deposits as part of the aging process and in response to various cardiovascular risk factors. ACs were historically considered an inevitable consequence of aging, but modern research has shown that these deposits result from active and biologically regulated processes that involve a complex interplay of cellular and molecular mechanisms, sharing similarities with atherosclerosis pathogenesis, including the deposition of lipids, the initiation of chronic inflammation, and the transformation of vascular smooth muscle cells into osteoblast-like cells capable of mineral deposition[4–6]. This revised understanding suggests that ACs are not merely benign markers of aging but rather manifestations of ongoing pathological processes that can potentially be influenced by targeted therapeutic interventions. Despite this, ACs are frequently under-reported, even with more holistic imaging modalities, like CT[7], and even when called, ACs are often not coded at myriad health systems, including the University of Texas Medical Branch at Galveston. These diagnosis and coding errors contribute to downstream problems with patient health and health system financial solvency.

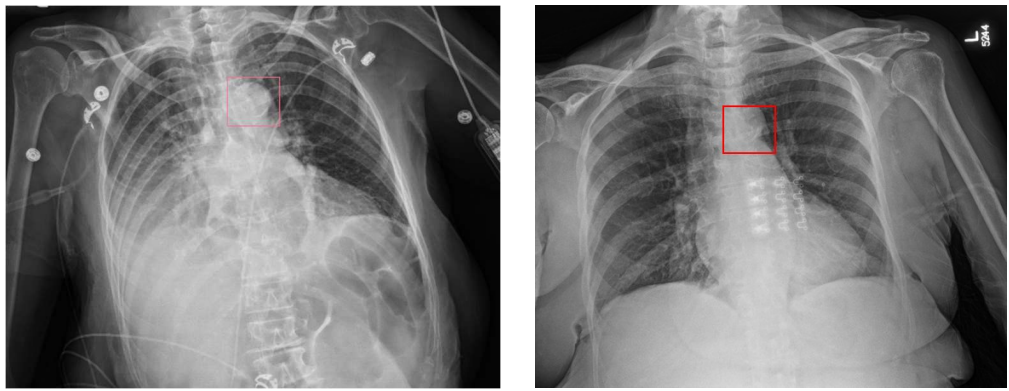

*Figure 1: Evident (left) and subtle (right) aortic calcifications identified using chest radiography for two de-identified patients. These calcifications are often overlooked, but provide highly valuable information to radiologists & referring providers regarding patient prognosis. Notably, the patient whose image is shown on the right subsequently had fatal complications from undiagnosed vascular disease.*

Chest radiography, a widely available, cost-effective, and frequently utilized imaging modality, plays a significant role in clinical practice for the evaluation of various thoracic conditions. Often performed as a routine examination or for specific indications such as respiratory symptoms, chest radiographs can be used to incidentally detect ACs[8,9]. The high prevalence and accessibility of chest X-rays (CXRs) yield a valuable tool for opportunistic identification of ACs, potentially leading to earlier recognition of and treatment for underlying cardiovascular risks.

Beyond their association with atherosclerosis, ACs indicate increased risks for myriad cardiovascular diseases and events, including coronary artery disease, stroke, aortic stenosis, and overall cardiovascular mortality[2,10]. Therefore, early AC detection can prompt clinicians to further evaluate patients' cardiovascular health and implement appropriate preventive strategies[10]. The intent of this section is to comprehensively address the importance of AC detection for patient management and to detail the differences in patient treatment strategies for affected vs unaffected patients and to explore the prognostic implications associated with early versus late diagnosis.

## 1. Influence of Aortic Calcification Diagnosis on Treatment Strategies

Treatment strategies for patients diagnosed with aortic calcifications often differ significantly from those without this finding[12]. The presence of these calcium deposits indicates an underlying atherosclerotic process and necessitates a focused approach to manage this condition and its potential consequences[11].

For medical therapy, aggressive risk factor modification becomes paramount in patients with ACs. This includes comprehensive lifestyle interventions, including dietary changes to reduce cholesterol and saturated fat intake, increased physical exercise to improve cardiovascular health, and smoking cessation. Pharmacological management may also be provided, for example via prescription of statins to lower lipid levels and potentially slow the progression of atherosclerosis and aortic stenosis[13]. Antihypertensive medications are essential for controlling blood pressure[11], and in patients with diabetes, strict glycemic control is crucial[14]. The diagnosis of aortic calcification can serve as a powerful motivator for both patients and clinicians to adopt or intensify these risk-reducing behaviors and therapies.

## 2. Prognostic Implications of Early Versus Late Diagnosis of Aortic Calcification

The timing of diagnosis for aortic calcification carries significant prognostic implications. Patients for whom ACs are detected early may experience different long-term outcomes compared to those diagnosed later, often when symptoms have already manifested (give examples)[16]. Early diagnosis provides a valuable opportunity for timely implementation of primary and secondary prevention strategies, which can potentially slow the progression of vascular disease and improve overall cardiovascular outcomes[15].

Early detection of aortic arch calcification via CXRs has been consistently shown to be a strong independent predictor of future cardiovascular events, including stroke, myocardial infarction, and cardiovascular death[8]. This predictive power often extends beyond that of traditional risk factors alone, suggesting that earlier identification through this readily available imaging modality can prompt timely interventions and more proactive risk factor management.

## Section A conclusion

The detection of aortic calcification using chest radiographs holds significant value for patient health management as a crucial marker of underlying cardiovascular disease with substantial prognostic value. The presence of ACs, even in asymptomatic individuals, signifies an increased risk of future cardiovascular events and warrants consideration in clinical decision-making.

Treatment strategies for patients with ACs often involve a more intensive approach to managing traditional cardiovascular risk factors through lifestyle modifications and pharmacological therapies. Opportunistic detection of aortic calcification through chest radiography has the potential to positively impact patient outcomes by enabling timely interventions and aggressive risk factor management.

## Section B: Financial implications of aortic calcification detections

The detection of aortic calcification, often incidentally on chest radiographs, serves as a critical marker for underlying cardiovascular disease and initiates a series of diagnostic and therapeutic actions that significantly impact healthcare finances. The University of Texas Medical Branch at Galveston performed an actuarial assessment that showed that each patient identified & coded with aortic calcifications resulted in an average annual value of $542 per patient to the health institution. This annual value reflects the balance between the increased revenue from diagnostic and treatment procedures and the associated costs and is a blended average across all patient populations for patients above 50 years of age, regardless of insurance, reimbursement, or risk adjustment mechanism. Early detection can potentially prevent more expensive complications, contributing to long-term cost savings. This value delivery to healthcare systems like UTMB is critical to enable these systems to remain financially viable and continue providing patient care.

## Section C: Discovery efforts in collaboration with Zauron Labs, Inc.

Given the clinical & financial importance of identifying patients with aortic calcifications, UTMB partnered with Zauron Labs, Inc., a Radiology Imaging Quality & Safety company, to perform a retrospective quality assessment to help identify patients with missed or uncoded aortic calcifications. This discovery effort comprised two separate thrusts: report plus imaging assessments to identify patients whose ACs were not diagnosed and exclusively report-based assessments to discover correct diagnoses that were not appropriately billed. Section C provides a summary of the results obtained by Zauron from this preliminary pilot effort, along with a plan for identifying and treating the remainder of UTMB's patients.

### 1. Exclusion of patients with known ACs

Prior to execution of any discovery efforts, UTMB removed any patients whose medical records already reflected the presence of aortic calcifications. UTMB only considered adult patients whose exams were reviewed within one year prior to the assessment, who underwent imaging in the form of chest X-rays, and who did not have any pre-existing records of ACs. A random selection of 3,988 patient exams (total of 5000 image, report pairs) from this cohort was selected for inclusion.

### 2. Language model based detections of diagnosed, uncoded Aortic Calcifications

Radiologists use a variety of terms to indicate the presence of aortic calcifications, for example, "Atherosclerotic calcifications of the aorta" or "Calcifications in the aortic arch". Additionally, the voice to text software used by many radiologists during exam dictations can erroneously yield nonsensical text, including "Constipation [sic] are seen in the aortic arch". This broad nomenclature and potential for translation errors can result in high levels of confusion and, more importantly, coding errors among teams tasked with submitting claims to receive reimbursements for treatments that hospitals provide to patients with ACs. The recent proliferation of language models provides an opportunity to identify these patient exams and correct claims to protect UTMB's solvency.

Zauron's proprietary language model is a local deployment of Meta's Llama 3.2 Open Source LLM. From the aforementioned cohort of 3,988 patients, Zauron identified 495 patients whose report contained aortic calcification diagnoses that were not properly coded for reimbursement, translating to a coding or documentation miss rate of ~**12.4%**.

## 3. Computer vision-based detections of non-diagnosed, uncoded Aortic Calcifications

The most important patients in the quality assessment Zauron performed for UTMB were those whose initial exams had no known vascular disease by medical history and had no mention of vascular disease in their exam report. Zauron deployed its research use only aortic calcification detection algorithm (AUC = 0.938) to all exams in the sample and rank ordered the exams from most to least likely to have ACs. Exams and patients which were detected via the prior LLM analysis were then excluded (n=495 patients). Dr. Clark manually reviewed all remaining 101 patient's exams with non-zero confidences. Using this method, **84 new patients were identified** whose original prospective radiology reports did not include any indication of ACs, which translates into a minimum diagnostic miss rate of ~2.1%. Limited scope reports were issued for all reports to formally update their patient records and enable treatment and monitoring.

## 4. Summary of detections

As a result of the quality assessment Zauron performed for UTMB, UTMB was able to identify 84 patients who needed clinical follow ups as well as an additional 495 patients who received a diagnosis which was never billed or coded. Mitigation of these clinical and financial risks helps protect patient health and institutional solvency to continue providing top notch care to patients.

**Section D: Closing the gap - How UTMB is deploying Zauron technology to avoid misdiagnoses moving forward**

Given the success of this preliminary pilot, UTMB is now leveraging Zauron's quality assessment services for the remainder of all eligible patients to ensure that equal treatment is provided to all patients. As a world class institution, it is critical to UTMB that diagnoses are not missed that could negatively impact patient health. Further, accreditation bodies like the American College of Radiology require all accredited radiology practices to perform regular quality measures either in the form of Peer Review or Peer Learning.

Legacy Peer Review software programs, like RADPEER or PACS provider integrated versions of RADPEER, have been ineffective at identifying discrepancies or misdiagnoses, but are still heavily used by radiologists to meet accreditation requirements. Zauron's Guardian Pro software retrospectively deploys a suite of image algorithms to identify abnormalities in images, along with a language model to identify what abnormalities were already mentioned during the original radiologist's prospective review. Any discrepancies for which image models indicate the presence of an abnormality that was not mentioned by the radiologist are selected for review as part of AI-enhanced Peer Review exercises. Radiologists can review findings to confirm or reject the presence of the identified abnormality and to further indicate the severity of the diagnosis using the streamlined graphical user interface shown below in Figure 2.

In addition to providing UTMB with discrepant reviews, Zauron's Guardian Pro software enables UTMB to easily organize Peer Learning conferences to focus on those abnormalities that were missed, whether identified by Guardian Pro or from random selection. This combination of Peer Learning & Peer Review helps UTMB's radiology practice and hospital system to maintain its status as a world class institution across the board.

UTMB has invested in patient safety by purchasing a five-year subscription to Zauron's Guardian Pro software. Zauron's Guardian Pro software is exclusively focused on patient health through the identification of discrepant exams that contain misdiagnoses. Record reviews for financial solvency are available to UTMB on an ad hoc basis from Zauron.

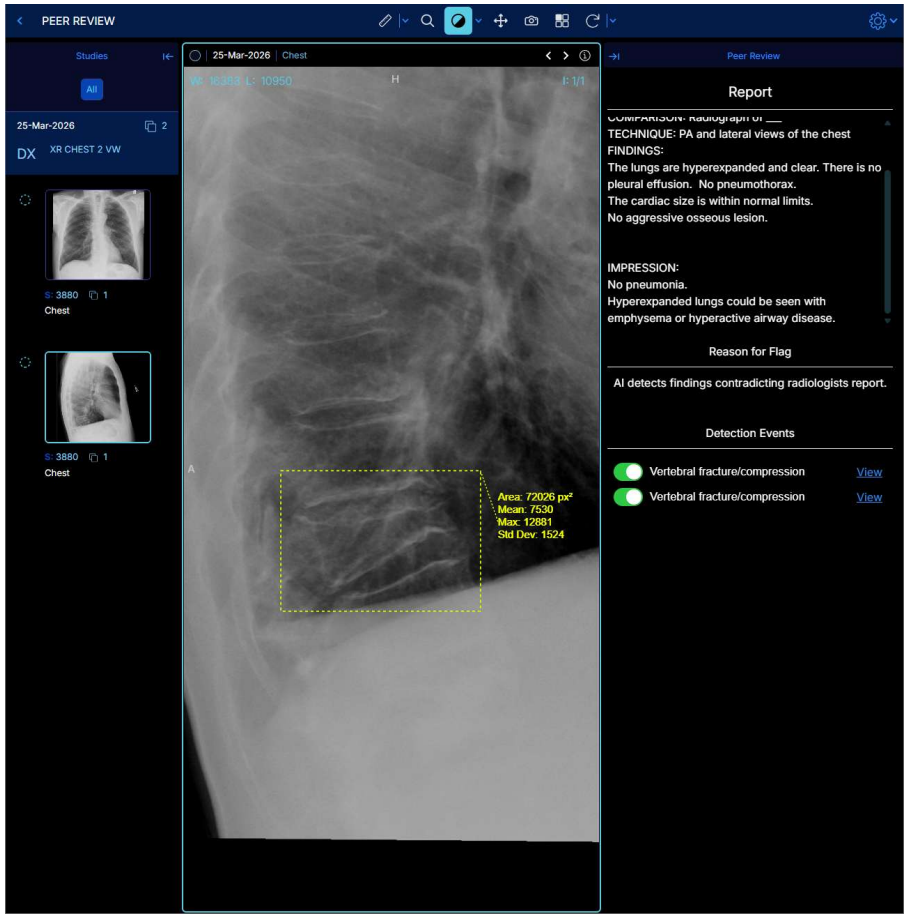

*Figure 2: Zauron Labs, Inc.'s Guardian Pro Peer Review software highlights abnormalities that were overlooked during preliminary reviews for second looks. This de-identified image shows an example of a vertebral compression fractures that was not mentioned by the original radiologist reviewing this exam. A UTMB radiologist was assigned this case by Zauron's Guardian Pro software for Peer Review quality exercises and confirmed the presence of this missed finding, allowing this patient to receive follow up care.*

1. Allison MA, Criqui MH, Wright CM. Patterns and Risk Factors for Systemic Calcified Atherosclerosis. *ATVB*. 2004;24(2):331-336. doi:10.1161/01.ATV.0000110786.02097.0c

2. Eisen A, Tenenbaum A, Koren-Morag N, et al. Calcification of the Thoracic Aorta as Detected by Spiral Computed Tomography Among Stable Angina Pectoris Patients: Association With Cardiovascular Events and Death. *Circulation*. 2008;118(13):1328-1334. doi:10.1161/CIRCULATIONAHA.107.712141

3. Allison MA, Cheung P, Criqui MH, Langer RD, Wright CM. Mitral and Aortic Annular Calcification Are Highly Associated With Systemic Calcified Atherosclerosis. *Circulation*. 2006;113(6):861-866. doi:10.1161/CIRCULATIONAHA.105.552844

4. O'Brien KD. Pathogenesis of Calcific Aortic Valve Disease: A Disease Process Comes of Age (and a Good Deal More). *ATVB*. 2006;26(8):1721-1728. doi:10.1161/01.ATV.0000227513.13697.ac

5. Pawade T, Sheth T, Guzzetti E, Dweck MR, Clavel MA. Why and How to Measure Aortic Valve Calcification in Patients With Aortic Stenosis. *JACC: Cardiovascular Imaging*. 2019;12(9):1835-1848. doi:10.1016/j.jcmg.2019.01.045

6. New SEP, Aikawa E. Molecular Imaging Insights Into Early Inflammatory Stages of Arterial and Aortic Valve Calcification. Towler DA, ed. *Circulation Research*. 2011;108(11):1381-1391. doi:10.1161/CIRCRESAHA.110.234146

7. Williams MC, Weir-McCall J, Moss AJ, et al. Radiologist opinions regarding reporting incidental coronary and cardiac calcification on thoracic CT. *BJR|Open*. 2022;4(1):20210057. doi:10.1259/bjro.20210057

8. Woo JS, Kim W, Kwon SH, et al. Aortic arch calcification on chest X-ray combined with coronary calcium score show additional benefit for diagnosis and outcome in patients with angina.

9. Chao CT, Yeh HY, Hung KY. Chest radiography deep radiomics-enabled aortic arch calcification interpretation across different populations. *iScience*. 2023;26(4):106429. doi:10.1016/j.isci.2023.106429

10. Leow K, Szulc P, Schousboe JT, et al. Prognostic Value of Abdominal Aortic Calcification: A Systematic Review and Meta-Analysis of Observational Studies. *JAHA*. 2021;10(2):e017205. doi:10.1161/JAHA.120.017205

11. Desai MY, Cremer PC, Schoenhagen P. Thoracic Aortic Calcification. *JACC: Cardiovascular Imaging*. 2018;11(7):1012-1026. doi:10.1016/j.jcmg.2018.03.023

12. Kramer B, Vekstein AM, Bishop PD, et al. Choosing transcatheter aortic valve replacement in porcelain aorta: outcomes versus surgical replacement. *European Journal of Cardio-Thoracic Surgery*. 2023;63(5):ezad057. doi:10.1093/ejcts/ezad057

13.	Chan KL, Teo K, Dumesnil JG, Ni A, Tam J. Effect of Lipid Lowering With Rosuvastatin on Progression of Aortic Stenosis: Results of the Aortic Stenosis Progression Observation: Measuring Effects of Rosuvastatin (ASTRONOMER) Trial. *Circulation*. 2010;121(2):306-314. doi:10.1161/CIRCULATIONAHA.109.900027

14.	Murali S, Smith ER, Tiong MK, Tan S, Toussaint ND. Interventions to Attenuate Cardiovascular Calcification Progression: A Systematic Review of Randomized Clinical Trials. *JAHA*. 2023;12(23):e031676. doi:10.1161/JAHA.123.031676

15.	Rajamannan NM, Otto CM. Targeted Therapy to Prevent Progression of Calcific Aortic Stenosis. *Circulation*. 2004;110(10):1180-1182. doi:10.1161/01.CIR.0000140722.85490.EA

16.	Freeman RV, Otto CM. Spectrum of Calcific Aortic Valve Disease: Pathogenesis, Disease Progression, and Treatment Strategies. *Circulation*. 2005;111(24):3316-3326. doi:10.1161/CIRCULATIONAHA.104.486738

17.	Lindman BR, Clavel MA, Mathieu P, et al. Calcific aortic stenosis. *Nat Rev Dis Primers*. 2016;2(1):16006. doi:10.1038/nrdp.2016.6

18.	Lantelme P, Eltchaninoff H, Rabilloud M, et al. Development of a Risk Score Based on Aortic Calcification to Predict 1-Year Mortality After Transcatheter Aortic Valve Replacement. *JACC: Cardiovascular Imaging*. 2019;12(1):123-132. doi:10.1016/j.jcmg.2018.03.018